\documentclass[a4paper,11pt]{article}  

\usepackage{jheppub1}

\usepackage[T1]{fontenc} 
\usepackage{float}
\usepackage{epstopdf}

\usepackage{hyperref} 
\usepackage{cleveref}

\title{\boldmath Greybody factor of scalar fields from black strings}

\author[a,b]{Jamil Ahmed} 
\author[a,c]{and K. Saifullah}

\affiliation [a]{Department of Mathematics, Quaid-i-Azam University, Islamabad, Pakistan}
\affiliation [b]{Department of Physics and Astronomy, University of Waterloo, Waterloo, ON, N2L 3G1, Canada}
\affiliation[c]{Center for the Fundamental Laws of Nature, Harvard University, Cambridge, MA 02138, USA \\ }

 \affiliation{\emph{Electronic Address: jahmed@student.qau.edu.pk; ksaifullah@fas.harvard.edu}}

\abstract{The greybody factor of massless, uncharged scalar fields is studied in
the background of cylindrically symmetric spacetimes, in the low-energy
approximation. We discuss two cases. In the first case we derive
analytical expression for the absorption probability when the spacetime is
kinetically coupled with the Einstein tensor. In the second case we do the analysis
in the absence of the coupling constant. For this purpose we analyze the wave equation which
is obtained from Klein-Gordon equation. The radial part of the wave equation is solved
in the form of the hypergeometric function in the near horizon region, whereas in
the far region the solution is of the form of Bessel's function. Finally,
considering continuity of the wave function we smoothly match the two
solutions in the low energy approximation to get the formula for the absorption
probability.}

\begin{document}
\maketitle
\flushbottom

\section{Introduction}
\label{sec:intro}

Black holes are the most interesting objects worth investigating in any
gravitational theory. Considering black holes as thermal systems their entropy  and thermodynamics were investigated by taking into account quantum
mechanical effects \cite{JBS,J}. Thus black holes have an associated temperature and entropy
and therefore they radiate, and the radiations are called Hawking
radiations. Hawking temperature of radiations
emitted from different black holes has been studied \cite{UK, MK, AK1}. 
The emission rate at the event horizon of a black hole, in a mode with frequency $\omega$, is given by \cite{SW}
\begin{equation}
\Gamma (\omega )=\left( \frac{1d^{3}k}{e^{\beta \omega }\pm 1(2\pi )^{3}}\right) ,  \label{1}
\end{equation}
where $\beta$ is the inverse of Hawking temperature and the minus (plus) sign is
for bosons (fermions). This formula for the emission rate can be generalized for
any dimension and it is valid for massive and massless particles. Therefore
at the event horizon the spectrum of the radiations from black holes is
perfectly the same as that of the black-body spectrum. Due to this it gives rise
to the information loss paradox. The important fact is that geometry of the
spacetime around a black is non-trivial. This non-trivial geometry modifies
the spectrum of Hawking radiations. In fact the non-trivial geometry acts
as a potential barrier which allows some of the radiations to transmit and
to reflect the rest to the hole. 

The greybody factor, defined as the probability for a given
wave coming from infinity to be absorbed by the black hole (rate of absorption probability), is directly connected to the absorption cross section \cite{YH, TJ, JAKS, JAS, AFB, SI, IS, MF, WJ, L1}. The mathematical expression that summarizes the above discussion is
\begin{equation}
\Gamma (\omega )=\left( \frac{\gamma (\omega )d^{3}k}{e^{\beta \omega }\pm
1(2\pi )^{3}}\right) ,  \label{2}
\end{equation}
where $\gamma (\omega)$ is the so-called greybody factor, which is
frequency dependent.

Physically the greybody factor originates from an effective potential barrier
by a black hole spacetime. For example the potential barrier for massless
scalars from Schwarzschild spacetime is
\begin{equation}
V_{eff}\left( x\right) =\left( 1-\frac{r_{H}}{r}\right) \left( \frac{r_{H}}{r^{3}}+\frac{l(l+1)}{r^{2}}\right) ,  \label{3}
\end{equation}
with the tortoise coordinate
\begin{equation}
x=r+r_{H}\ln \left( \frac{r}{r_{H}}-1\right) ,  \label{4}
\end{equation}
where $r_{H}$ is the horizon radius and $l$ is the angular momentum of the
scalar. It is this potential which transmits or reflects radiations from black
holes. Therefore it gives rise to the frequency dependent greybody factor.
This factor not only accounts for the deviation of Hawking radiations
from the black-body spectrum, but it also could be important in the energy emission
rate and relevant to compute the partial absorption cross section of
black holes. The main idea to obtain the expression for the greybody factor is to
derive the solution of the relevant wave equation in near horizon and asymptotic regions
separately and then match them to an appropriate intermediate point \cite{JAS, AFB, SI, IS, MF, WJ, D, W, PJA}.

Scalar fields, non-minimally coupled with gravity, have shown significant
features, both for inflation and dark energy. Also, the non-minimal
couplings between derivatives of the scalar fields and the curvature
reveal interesting cosmological behaviours. In general, the scalar-tensor
theories give both the Einstein equation and the equation of motion for the
scalar in the form of fourth-order differential equations. But if the
kinetic term is only coupled to the Einstein tensor, the equation of motion for
scalars is reduced to a second-order differential equation. Therefore, from the
point of view of physics, considering such a coupling can be interpreted as
a good theory because it is very simple. In the light of the earlier results \cite
{G11, G12, SS11} there is a need for more efforts to be focussed on the study of scalar fields coupled with tensors for more general cases. In order to
fill the gap in the literature, for the case of cylindrically symmetric black
holes, we have studied the properties of the scalar field when it is
kinetically coupled to the Einstein tensor and the one without any coupling,
separately.

The rest of this letter is organized as follows. In Section 2 Klein-Gordon equation in
a charged black string background with coupling to the Einstein tensor is given.
In Section 3 solutions of the radial equation resulting from the Klein-Gordon
equation in the near horizon region and the far horizon regime are presented. These are also matched to an intermediate region to get a value of the absorption
probability. In Section 4 we do the above analysis
in the absence of the coupling parameter. Section 5 gives some concluding remarks.


\section{Klein-Gordon equation in the background of charged black string}
The Klein Gordon equation when the Einstein tensor is coupled to a massless, uncharged
scalar field is
\begin{equation}
\frac{1}{\sqrt{-g}}\partial _{\mu } \Big[ \sqrt{-g}\left( g^{\mu \nu }+\eta
\epsilon ^{\mu \nu }\right) \partial _{\nu }\Psi \Big] =0,  \label{9}
\end{equation}
where $\eta$ is a coupling constant and $\epsilon ^{\mu \nu}$ is Einstein's tensor. The charged black string having non-zero components of Einstein's tensor is
\cite{JK}
\begin{equation}
ds^{2}=-f(r)dt^{2}+\frac{1}{f(r)}dr^{2}+r^{2}d\theta ^{2}+\alpha
^{2}r^{2}dz^{2},  \label{10}
\end{equation}
where
\begin{equation}
f(r)=\alpha ^{2}r^{2}-\frac{4M}{\alpha r}+\frac{4Q^{2}}{\alpha ^{2}r^{2}}.
\label{11}
\end{equation}
Here $M$ is the mass, $Q$ is the charge and $\alpha =-\Lambda /3$, with $\Lambda$
being the cosmological constant. For the above metric the Einstein tensor $\epsilon ^{\mu \nu }$ in matrix form can be written as 
\begin{equation}
\epsilon ^{\mu \nu }=\frac{4Q^{2}}{\alpha ^{4}r^{4}}\left(
\begin{array}{cccc}
-\frac{1}{f} & 0 & 0 & 0 \\
0 & f & 0 & 0 \\
0 & 0 & -\frac{1}{r^{2}} & 0 \\
0 & 0 & 0 & -\frac{1}{\alpha ^{2}r^{2}}
\end{array}
\right) .  \label{12}
\end{equation}
Also 
\begin{equation}
\sqrt{-g}=\alpha r^{2}.  \label{14}
\end{equation}
Substituting the components of the Einstein tensor and spacetime metric in equation
(\ref{9}), it takes the form
\begin{eqnarray}
\frac{1}{\alpha r^{2}}\partial _{t}\bigg[ \alpha r^{2}\left( -\frac{1}{f}-
\frac{4\eta Q^{2}}{\alpha ^{4}r^{4}f}\right) \partial _{t}\Psi \bigg] +
\frac{1}{\alpha r^{2}}\partial _{r}\bigg[ \alpha r^{2}\left( f+\frac{4\eta
Q^{2}f}{\alpha ^{4}r^{4}}\right) \partial _{r}\Psi \bigg]+  \nonumber \\ \frac{1}{\alpha
r^{2}}\partial _{\theta }\bigg[ \alpha r^{2}\left( \frac{1}{r^{2}}-\frac{
4\eta Q^{2}}{\alpha ^{4}r^{6}}\right) \partial _{\theta }\Psi \bigg]
+\frac{1}{\alpha r^{2}}\partial _{z}\bigg[ \alpha r^{2}\left( \frac{1}{
\alpha r^{2}}-\frac{4\eta Q^{2}}{\alpha ^{6}r^{6}}\right) \partial _{\theta
}\Psi \bigg] =0.  \label{15}
\end{eqnarray}
Using the form of cylindrical harmonics
\begin{equation}
\Psi (t,r,\theta ,z)=e^{-\iota \omega t}R(r)Y(\theta ,z),  \label{16}
\end{equation}
we get from the radial part of equation $\left( \ref{15}\right)$ as
\begin{equation}
\frac{1}{r^{2}}\frac{d}{dr}\bigg[r^{2}\left( 1+\frac{4\eta
Q^{2}}{\alpha ^{4}r^{4}}\right) f\bigg] \frac{dR(r)}{dr}+\left[ \left( 1+%
\frac{4\eta Q^{2}}{\alpha ^{4}r^{4}}\right) \frac{\omega ^{2}}{f}-\left( 1-%
\frac{4\eta Q^{2}}{\alpha ^{4}r^{4}}\right) \frac{F_{lm}}{\alpha ^{2}r^{2}}%
\right] R(r)=0,  \label{17}
\end{equation}
where $F_{lm}=l(l+1)$ are the eigenvalues coming from the $(\theta ,z)$ part.


\section{Greybody factor computation}

\subsection{Near horizon solution}

Equation\ $\left( \ref{17}\right)$ is the master equation of our interest.
We will solve this equation in two regions separately, namely, the near
horizon region and the far region by using a semi-classical approach known as
the simple matching technique. We will match both solutions to an
intermediate region to get the analytical expression for the absorption
probability.

For the near horizon region $r\sim r_{+}$, we will perform the following
transformation to simplify the radial equation \cite{SOPK, DKS}: 
\begin{equation}
r\rightarrow f,  \label{18}
\end{equation}
which implies
\begin{equation}
\frac{df}{dr}=\left( 1-f\right) \frac{B(r_{+})}{r_{+}},  \label{19}
\end{equation}
where $r_+$ is the horizon and 
\begin{equation}
B(r_{+})=1-\frac{4Q^{2}-2\alpha ^{4}r_{+}^{4}}{4M\alpha r_{+}-4Q^{2}}.
\label{20}
\end{equation}
Using the above, equation $\left( \ref{17}\right) $ takes the form
\begin{eqnarray}
f(1-f)\frac{d^{2}R(f)}{df^{2}}+\left( 1-C_{\ast }f\right) \frac{dR(f)}{df} \nonumber \\ +
\left[ \frac{F_{\ast }^{2}}{B^{2}(r_{+})\left( 1-f\right) f}-\left( \frac{
\alpha ^{4}r_{+}^{4}-4\eta Q^{2}}{\alpha ^{4}r_{+}^{4}+4\eta Q^{2}}\right)
\frac{F_{lm}}{B^{2}(r_{+})\alpha ^{2}\left( 1-f\right) }\right] R(f)=0.
\label{21}
\end{eqnarray}
Here
\begin{equation}
F_{\ast }=\omega r_{\ast },  \label{22}
\end{equation}
and
\begin{equation}
C_{\ast }=2-\frac{2}{B(r_{+})}\left( \frac{\alpha ^{4}r_{+}^{4}-4\eta Q^{2}}{
\alpha ^{4}r_{+}^{4}+4\eta Q^{2}}\right) .  \label{23}
\end{equation}
In order to further simplify the above equation, we use the field redefinition
\begin{equation}
R(f)=f^{\mu }\left( 1-f\right) ^{\nu }F(f).  \label{24}
\end{equation}
Using this in equation $\left( \ref{21}\right)$, we obtain 

\begin{eqnarray}
f\left( 1-f\right) \frac{d^{2}F(f)}{df^{2}}+\left[ 1+2\mu -\left( 2\mu +2\nu
+C_{\ast }\right) f\right] \frac{dF}{df} \nonumber \\ + [\frac{\mu ^{2}}{f}-\mu ^{2}+\mu
-2\mu \nu +\frac{\nu ^{2}}{1-f}-\nu ^{2}-\frac{2\nu }{1-f}
+\nu -\mu C_{\ast }+\frac{\nu C_{\ast }}{1-f}-\nu C_{\ast }  \nonumber \\ +\frac{F_{\ast
}^{2}}{B^{2}(r_{+})f}+\frac{F_{\ast }^{2}}{B^{2}(r_{+})\left( 1-f\right)}
-\left( \frac{\alpha ^{4}r_{+}^{4}-4\eta Q^{2}}{\alpha ^{4}r_{+}^{4}+4\eta
Q^{2}}\right) \frac{F_{lm}}{B^{2}(r_{+})\alpha ^{2}\left( 1-f\right)}]F(f)=0.  \label{25}
\end{eqnarray}

We define
\begin{equation}
a=\mu +\nu +C_{\ast }-1, \,\,\ b=\mu +\nu , \,\,\ c=1+2\mu .  \label{26}
\end{equation}
Also the constraints coming from the coefficients of $F(f)$ give 

\begin{equation}
\mu ^{2}+\frac{F_{\ast }^{2}}{B^{2}(r_{+})}=0,  \label{27}
\end{equation}
and
\begin{equation}
\nu ^{2}+\nu (C_{\ast }-2)+\frac{F_{\ast }^{2}}{B^{2}(r_{+})}-\left( \frac{
\alpha ^{4}r_{+}^{4}-4\eta Q^{2}}{\alpha ^{4}r_{+}^{4}+4\eta Q^{2}}\right)
\frac{F_{lm}}{B^{2}(r_{+})\alpha ^{2}}=0.  \label{28}
\end{equation}
From this we get the values of $\mu$ and $\nu$: 
\begin{equation}
\mu _{\pm }=\pm \iota \frac{F_{\ast }}{B(r_{+})},  \label{29}
\end{equation}
and
\begin{equation}
\nu _{\pm }=\frac{1}{2}\left[ \left( 2-C_{\ast }\right) \pm \sqrt{\left(
2-C_{\ast }\right) ^{2}-4\left( \frac{F_{\ast }^{2}}{B^{2}(r_{+})}-\left(
\frac{\alpha ^{4}r_{+}^{4}-4\eta Q^{2}}{\alpha ^{4}r_{+}^{4}+4\eta Q^{2}}
\right) \frac{F_{lm}}{B^{2}(r_{+})\alpha ^{2}}\right) }\right] .  \label{30}
\end{equation}
Equation $\left( \ref{25}\right) $ by virtue of $\left( \ref{26}
\right) $ and the constraints $\left( \ref{27}\right) $-$\left( \ref{28}\right)$
becomes
\begin{equation}
f\left( 1-f\right) \frac{d^{2}F(f)}{df^{2}}+\bigg[ c-\left( 1+a+b\right) f
\bigg] \frac{dF(f)}{df}-abF(f)=0.  \label{31}
\end{equation}

For the near horizon case there exists no outgoing mode, which means  $\mu
_{+}=\mu _{-}$ and $\nu _{+}=\nu _{-}$. So in the near horizon region the
solution can be written in the form of the general hypergeometric function,
which has the form
\begin{equation}
R(f)_{NH}=C_{-}f^{\mu }(1-f)^{\nu }F\left( a,b,c;f\right) ,  \label{32}
\end{equation}
where $C_{-\text{ }}$ is an arbitrary constant. 


\subsection{Far horizon solution}
Now we find the solution of the radial equation for the far
region. In this case the radial part will have the form
\begin{equation}
\frac{d^{2}R(r)_{FR}}{dr^{2}}+\frac{4}{r}\frac{dR(r)_{FR}}{dr}+\left( \omega
^{2}-\frac{F_{lm}}{\alpha ^{2}r^{2}}\right) R(r)_{FR}=0.  \label{38}
\end{equation}
This is the well-known Bessel equation, and in a far field its solution can be
written as
\begin{equation}
R_{FR}\left( r\right) =\frac{1}{\sqrt{r\alpha \omega }}\left[ B_{1}J_{\gamma
}(\omega \alpha r)+B_{2}Y_{\gamma }(\omega \alpha r)\right] .  \label{39}
\end{equation}
In the above solution $J_{\gamma }$ and $Y_{\gamma }$ are Bessel's functions. For $\gamma =l+1/2$, and in the limit $r\rightarrow 0$, the above solution can
be written as 
\begin{equation}
R_{FR}\left( r\right) \simeq \frac{B_{1}\left( \frac{\omega \alpha r}{4}
\right) ^{\gamma }}{\sqrt{\omega \alpha r}\Gamma \left( \nu +1\right) }-
\frac{B_{2}\Gamma \left( \gamma \right) }{\pi \sqrt{\omega \alpha r}\left(
\frac{\omega \alpha r}{4}\right) ^{\nu }}.  \label{40}
\end{equation} 


\subsection {Matching the two solutions}
We now stretch the near horizon solution to an intermediate region \cite{SJ, MI} which gives
\begin{eqnarray}
R(f)_{NH}=C_{-}f^{\mu }\left( 1-f\right) ^{\nu }\bigg[\frac{\Gamma (c)\Gamma
(c-a-b)}{\Gamma (c-a)\Gamma (c-b)}F\left( a,b,a+b-c+1;1-f\right)  \nonumber \\ 
+\left( 1-f\right) ^{c-a-b}\frac{\Gamma (c)\Gamma (a+b-c)}{\Gamma (a)\Gamma
(b)}F\left( c-a,c-b,c-a-b+1;1-f\right) \bigg].  \label{33}
\end{eqnarray}
We can approximate $1-f$ for the case $r\gg r_{+}$ as
\begin{equation}
1-f\simeq \frac{4M}{\alpha r}.  \label{34}
\end{equation}
So, the form of the final solution for the near horizon case becomes 
\begin{equation}
R(r)_{NH}\simeq A_{1}r^{\nu }+A_{2}r^{-(\nu +C_{\ast }-2)}.  \label{35}
\end{equation}
Here we have chosen
\begin{equation}
A_{1}=C_{-}\left( \frac{4M}{\alpha }\right) ^{\nu }\frac{\Gamma (c)\Gamma
(c-a-b)}{\Gamma (c-a)\Gamma (c-b)},  \label{36}
\end{equation}
and
\begin{equation}
A_{2}=C_{-}\left( \frac{4M}{\alpha }\right) ^{-(\nu +C_{\ast }-2)}\frac{
\Gamma (c)\Gamma (a+b-c)}{\Gamma (a)\Gamma (b)}.  \label{37}
\end{equation}
In the low-energy limit we can use the approximation
\begin{equation}
-\nu \simeq l+O(\omega ^{2}),  \label{41}
\end{equation}%
\begin{equation}
\nu +C_{\ast }-2\simeq -\left( l+1\right) +O(\omega ^{2}).  \label{41a}
\end{equation}%
From equations $\left( \ref{35}\right) $ and $\left( \ref{40}\right)$ matching the coefficients and eliminating $C_{-}$ give
\begin{equation}
B=\frac{B_{1}}{B_{2}}=-\frac{1}{\pi }\frac{1}{\left( \alpha \omega M\right)
^{2l+1}}\frac{\Gamma (c-a-b)\Gamma (a)\Gamma (b)}{\Gamma (c-a)\Gamma
(c-b)\Gamma (a)\Gamma (b)}\Gamma ^{2}(l+1/2).  \label{42a}
\end{equation}

The greybody factor can now be given by \cite{LAEJ}
\begin{equation}
\gamma _{l}\left( \omega \right) =\left\vert P_{l}\right\vert ^{2}=\frac{
2\iota \left( B^{\ast }-B\right) }{\left\vert B\right\vert ^{2}}.
\label{43}
\end{equation}
By using the value of $B$ we can find the expression of absorption
probability of the radiations emitted from the charged black string. This
relation gives a  measure of  how much the radiations are different (or
modified) from the spectrum of the black body radiation. 


\section {Absorption probability for scalar field without coupling to the Einstein tensor}
In this section we find an analytical expression of the absorption
probability for scalar field from the charged black string without coupling to
the Einstein tensor. The Klein-Gordon equation for a massless,
uncharged scalar field is 
\begin{equation}
\frac{1}{\sqrt{-g}}\partial _{\mu }\left[ \sqrt{-g}\left( g^{\mu \nu
}\right) \partial _{\nu }\Psi \right] =0.  \label{1a}
\end{equation}
Using the values of each component of the spacetime considered in the
previous section, we get the following equation: 
\begin{eqnarray}
\frac{1}{\alpha r^{2}}\partial _{t}\bigg[ \alpha r^{2}\left( -\frac{1}{f}
\right) \partial _{t}\Psi \bigg] +\frac{1}{\alpha r^{2}}\partial _{r}\bigg[
\alpha r^{2}\left( f\right) \partial _{r}\Psi \bigg] +  \nonumber \\ \frac{1}{\alpha r^{2}}
\partial _{\theta }\bigg[ \alpha r^{2}\left( \frac{1}{r^{2}}\right) \partial
_{\theta }\Psi \bigg] + \frac{1}{\alpha r^{2}}\partial _{z}\bigg[ \alpha
r^{2}\left( \frac{1}{\alpha r^{2}}\right) \partial _{\theta }\Psi \bigg] =0.
\label{2a}
\end{eqnarray}
Considering cylindrical harmonics, we can separate the radial part of
equation $\left( \ref{2a}\right)$, which has the form
\begin{equation}
\frac{1}{\alpha r^{2}}\frac{d}{dr}\left( \alpha r^{2}f\right) \frac{dR(r)}{dr
}+\left[ \frac{\omega ^{2}}{f}-\frac{F_{lm}}{\alpha ^{2}r^{2}}\right] R(r)=0.
\label{3a}
\end{equation} 



As in the previous case we will find two solutions of the radial
equation $\left( \ref{3a}\right)$, one for the near horizon and the other for
the far horizon regime. In the case of the near horizon region, we use the transformation $r\rightarrow f$, 
which implies
\begin{equation}
\frac{df}{dr}=\left( 1-f\right) \frac{B(r_{+})}{r_{+}},  \label{5a}
\end{equation}
where 
\begin{equation}
B(r_{+})=1-\frac{4Q^{2}-2\alpha ^{4}r_{+}^{4}}{4M\alpha r_{+}-4Q^{2}}.
\label{6a}
\end{equation}
Using equation $\left( \ref{5a}\right)$,
equation $(\ref{3a})$ takes the form
\begin{equation}
f(1-f)\frac{d^{2}R(f)}{df^{2}}+\left( 1-C_{\ast }f\right) \frac{dR(f)}{df}+
\left[ \frac{F_{\ast }^{2}}{B^{2}(r_{+})\left( 1-f\right) f}-\frac{F_{lm}}{
B^{2}(r_{+})\alpha ^{2}\left( 1-f\right) }\right] R(f)=0.  \label{7a}
\end{equation}
Here
\begin{equation}
F_{\ast }=\omega r_{\ast },  \,\,\,\ C_{\ast }=2-\frac{2}{B(r_{+})}.  \label{8a}
\end{equation}
In order to further simplify the above equation, we use field redefinition
\begin{equation}
R(f)=f^{\mu }\left( 1-f\right) ^{\nu }F(f).  \label{10a}
\end{equation}
In equation $\left( \ref{7a}\right)$ we use this definition of $R(f)$ to obtain 

\begin{eqnarray}
f\left( 1-f\right) \frac{d^{2}F(f)}{df^{2}}+\bigg[ 1+2\mu -\left( 2\mu +2\nu
+C_{\ast } \right) f \bigg] \frac{dF}{df}+ \nonumber \\ 
\bigg[ \frac{\mu ^{2}}{f}-\mu ^{2}+\mu
-2\mu \nu +\frac{\nu ^{2}}{1-f}-\nu ^{2}-\frac{2\nu }{1-f}+\nu -\mu C_{\ast }+\frac{\nu C_{\ast }}{1-f}-\nu C_{\ast }+  \nonumber \\ 
\frac{F_{\ast}^{2}}{B^{2}(r_{+})f}+\frac{F_{\ast }^{2}}{B^{2}(r_{+})\left( 1-f\right) }
-\left( \frac{\alpha ^{4}r_{+}^{4}-4\eta Q^{2}}{\alpha ^{4}r_{+}^{4}+4\eta
Q^{2}}\right) \frac{F_{lm}}{B^{2}(r_{+})\alpha ^{2}\left( 1-f\right) } 
\bigg]F(f)=0.  \label{11a}
\end{eqnarray}
We again use the definitions given in (\ref{26}). The constraints coming from the coefficients of $F(f)$ yield

\begin{equation}
\mu ^{2}+\frac{F_{\ast }^{2}}{B^{2}(r_{+})}=0,  \label{13a}
\end{equation}
and
\begin{equation}
\nu ^{2}+\nu (C_{\ast }-2)+\frac{F_{\ast }^{2}}{B^{2}(r_{+})}-\frac{F_{lm}}{
B^{2}(r_{+})\alpha ^{2}}=0.  \label{14a}
\end{equation}
These give the values of $\mu$ and $\nu$ as
\begin{equation}
\mu _{\pm }=\pm \iota \frac{F_{\ast }}{B(r_{+})},  \label{15a}
\end{equation}

\begin{equation}
\nu _{\pm }=\frac{1}{2}\left[ \left( 2-C_{\ast }\right) \pm \sqrt{\left(
2-C_{\ast }\right) ^{2}-4\left( \frac{F_{\ast }^{2}}{B^{2}(r_{+})}-\frac{
F_{lm}}{B^{2}(r_{+})\alpha ^{2}}\right) }\right] .  \label{16a}
\end{equation}
Equation $\left( \ref{11a}\right)$ by virtue of the above constraints becomes
\begin{equation}
f\left( 1-f\right) \frac{d^{2}F(f)}{df^{2}}+\bigg[ c-\left( 1+a+b\right) f
\bigg] \frac{dF(f)}{df}-abF(f)=0.  \label{17a}
\end{equation}

For the near horizon case there exists no outgoing mode, which means $\mu
_{+}=\mu _{-}$ and $\nu _{+}=\nu _{-}$. So, in the near horizon region the
solution can be written in the form of the general hypergeometric function, being of
the form
\begin{equation}
R(f)_{NH}=C_{1-}f^{\mu }(1-f)^{\nu }F\left( a,b,c;f\right) ,  \label{18a}
\end{equation}
where $C_{1-}$ is an arbitrary constant. We now stretch the near
horizon solution to an intermediate region \cite{SJ, MI} so that
\begin{eqnarray}
R(f)_{NH}=C_{-}f^{\mu }\left( 1-f\right) ^{\nu }\bigg[\frac{\Gamma (c)\Gamma
(c-a-b)}{\Gamma (c-a)\Gamma (c-b)}F\left( a,b,a+b-c+1;1-f\right)+ \nonumber \\ 
\left( 1-f\right) ^{c-a-b}\frac{\Gamma (c)\Gamma (a+b-c)}{\Gamma (a)\Gamma
(b)}F\left( c-a,c-b,c-a-b+1;1-f\right) \bigg].  \label{19a}
\end{eqnarray}
We again approximate $1-f$ for the case $r\gg r_{+}$, as before, and obtain the final form of the solution given in (\ref{35}). 

Now, as in the previous section, the radial equation for the far region reduces to the form of Bessel's equation, and the form of the final solution in this region is 
\begin{equation}
R_{FR}\left( r\right) \simeq \frac{B_{1}\left( \frac{\omega \alpha r}{4}
\right) ^{\gamma }}{\sqrt{\omega \alpha r}\Gamma \left( \nu +1\right) }-
\frac{B_{2}\Gamma \left( \gamma \right) }{\pi \sqrt{\omega \alpha r}\left(
\frac{\omega \alpha r}{4}\right) ^{\nu }}.  \label{26a}
\end{equation}%
Using the same procedure as in the previous case, we find
\begin{equation}
B=\frac{B_{1}}{B_{2}}=-\frac{1}{\pi }\frac{1}{\left( \alpha \omega M\right)
^{2l+1}}\frac{\Gamma (c-a-b)\Gamma (a)\Gamma (b)}{\Gamma (c-a)\Gamma
(c-b)\Gamma (a)\Gamma (b)}\Gamma ^{2}(l+1/2).  \label{28a}
\end{equation}

The absorption probability and hence greybody factor can be found by using
the value of $B$ in equation $\left( \ref{43}\right)$.
\begin{figure}[H]
  \centering
  \includegraphics[scale=1.2,]{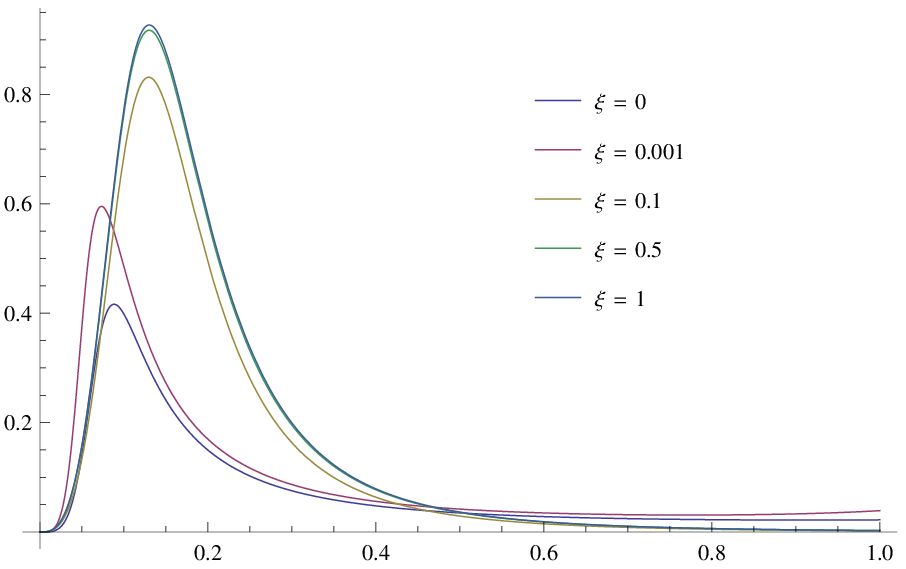}
  \caption{Greybody factor as a function of the frequency, for $\xi=0, 0.001, 0.1, 0.5, 1$ and $l=1$}\label{Greybody factor 2}
\end{figure}
\begin{figure}[H]
  \centering
  \includegraphics[scale=1.2,]{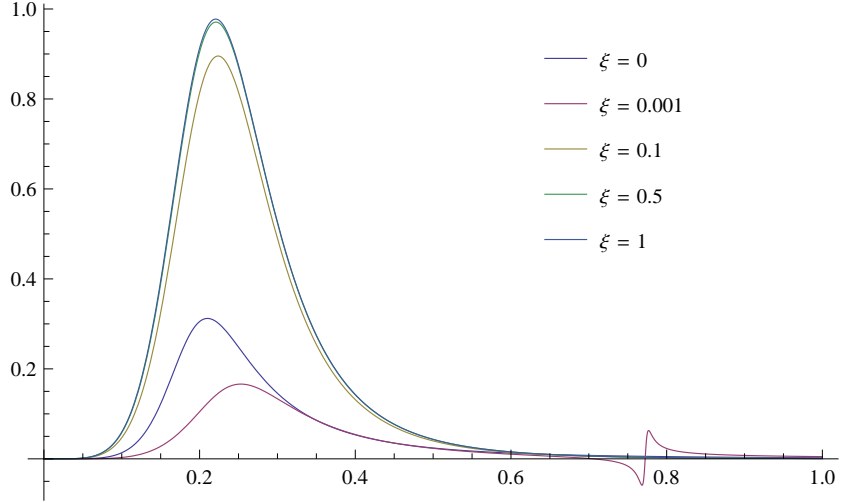}
  \caption{Greybody factor as a function of the frequency, for $\xi=0, 0.001, 0.1, 0.5, 1$ and $l=2$}\label{Greybody factor}
\end{figure}

The effect of the coupling constant on the greybody factor is also analyzed graphically for different partial modes. In Fig. 1, we draw the graph of the greybody factor as a function of the frequency for different values of the coupling constant and for $l=1$. In Fig. 2 it is depicted for $l=2$. It is observed that, for different modes, a stronger coupling enhances the absorption probability in the low frequency approximation. 


\section {Conclusion}

In this letter we have studied the greybody factor for a scalar field
coupling to the Einstein tensor in the background of a charged black string in
the low energy approximation. We found that the absorption probability and hence
the greybody factor depend on the coupling between the scalar field and the Einstein
tensor. It is observed that the presence of a coupling enhances the absorption
probability of the scalar field in the black string spacetime. Also, for
weaker coupling, the absorption probability decreases with the increase in the charge of black
string. In the second case we did this analysis without considering a coupling
of the scalar field and the Einstein tensor. Needless to say that the latter
case reduces to the result of the former in the absence of the coupling constant.

\acknowledgments
A research grant from the Higher Education Commission of Pakistan under its Project no. 20-2087 is gratefully acknowledged.

\end{document}